\documentclass{article}
\usepackage{ijcai16}

\usepackage{amsmath}
\usepackage{amsthm}
\usepackage{graphicx}
\usepackage{epstopdf}
\usepackage{subfigure}
\usepackage{multirow}
\usepackage{algorithmic}
\usepackage[ruled,vlined]{algorithm2e}
\usepackage{enumitem}
\setlist[itemize]{itemsep=-0.25ex,leftmargin=2.5ex}
\setlist[enumerate]{itemsep=-0.25ex,leftmargin=2.5ex}

\newcommand{\Sig}{\Theta}
\newcommand{\N}{\mathcal{N}}
\newcommand{\ie}{\emph{i.e.,}\xspace}
\newcommand{\eg}{\emph{e.g.,}\xspace}

\DeclareMathOperator*{\argmin}{arg\,min}

\usepackage{times}
\usepackage{amsmath,amssymb}

\title{Learning Optimal Social Dependency for Recommendation}

\author{
Yong Liu, Peilin Zhao, Xin Liu, Min Wu, Xiao-Li Li\\
Institute for Infocomm Research, A*STAR, Singapore\\
\{liuyo, zhaop, liu-x, wumin, xlli\}@i2r.a-star.edu.sg
}

\begin{document}

\maketitle

\begin{abstract}
Social recommender systems exploit users' social relationships to improve the recommendation accuracy. Intuitively, a user tends to trust different subsets of her social friends, regarding with different scenarios. Therefore, the main challenge of social recommendation is to exploit the optimal social dependency between users for a specific recommendation task. In this paper, we propose a novel recommendation method, named probabilistic relational matrix factorization (PRMF), which aims to learn the optimal social dependency between users to improve the recommendation accuracy, with or without users' social relationships. Specifically, in PRMF, the latent features of users are assumed to follow a matrix variate normal (MVN) distribution. The positive and negative dependency between users are modeled by the row precision matrix of the MVN distribution. Moreover, we have also proposed an efficient alternating algorithm to solve the optimization problem of PRMF. The experimental results on real datasets demonstrate that the proposed PRMF method outperforms state-of-the-art social recommendation approaches, in terms of root mean square error (RMSE) and mean absolute error (MAE).

\end{abstract}

\section{Introduction}

Recommender systems have been widely used in our lives to help us discover useful information from a large amount of data. For example, on the popular e-commerce websites such as Amazon, the recommender systems predict users' preferences on products based on their past purchasing behaviors and then recommend a user a list of interesting products she may prefer~\cite{linden2003amazon}. For a recommender system, the most critical factor is the prediction accuracy of users' preferences. In practice, the most successful prediction methods are collaborative filtering based approaches, especially the matrix factorization models~\cite{su2009survey}.

The recent rapid developments of online social networking services (\eg Facebook and Twitter) motivate the emergences of social recommender systems that exploit users' online social friendships for recommendation~\cite{yang2014survey}. The social recommender systems usually assume that a user has similar interests with her social friends.
Although the recommendation accuracy can usually be improved, there do not exist strong connections between the online friendship and the similarity of users' interests~\cite{ma2014measuring}. Because there are different categories of social networking friends in online social networks, \eg school friends, work-related friends, friends sharing same interest/activities, and neighborly friends~\cite{zhang2013categories}. The tastes of a user's online friends usually vary significantly. In different scenarios, a user tends to trust the recommendations from different subsets of her online social friends. Hence, the key to success for a social recommender system is to exploit the most appropriate social dependency between users for recommendation.

Moreover, existing social recommender systems are usually developed based on users' explicit social relationships, \eg trust relationships~\cite{jamali2009trustwalker,jamali2010matrix} and online social friendships~\cite{ma2011recommender,yang2012circle,tang2013exploiting}. They ignore the underlying implicit social relationships between users that have most similar or dissimilar rating behaviors. This implicit social relationships have been found to be beneficial for improving the recommendation accuracy~\cite{ma2013experimental}. In addition, users' explicit social relationships may be unavailable in many application scenarios. This also limits the application of traditional social recommendation approaches.

In previous studies, users' social dependency adopted for recommendation are predefined based on users' explicit or implicit social relationships~\cite{ma2013experimental,yang2014survey}. Differing from previous work, this paper proposes a novel social recommendation method, namely probabilistic relational matrix factorization (PRMF), which aims to learn the optimal social dependency between users to improve the recommendation accuracy. The proposed method can be applied to the recommendation scenarios with or without users' explicit social relationships. In PRMF, the user latent features are assumed to follow a matrix variate normal (MVN) distribution. The positive and negative social dependency between users are modeled by the row precision matrix of the MVN distribution. This is motivated by the success of using the MVN distribution to model the task relationships for multi-task learning~\cite{zhang2010convex}. To solve the optimization problem of PRMF, we propose a novel alternating algorithm based on the stochastic gradient descent (SGD)~\cite{koren2009matrix} and alternating direction method of multipliers (ADMM)~\cite{boyd2011distributed} methods. Moreover, we extensively evaluated the performances of PRMF on four public datasets. Empirical experiments showed that PRMF outperformed the state-of-the-art social recommendation methods, in terms of root-mean-square error (RMSE) and mean absolute error (MAE).

\section{Related Work and Background}
\label{section:relatedwork}
In this section, we first introduce some background about probabilistic matrix factorization (PMF)~\cite{mnih2007probabilistic}, one of the most popular matrix factorization models. Then, we review the state-of-the-art social recommendation methods.

\subsection{Probabilistic Matrix Factorization}
\label{ssection:pmf}
For the recommendation problem with $m$ users $\{u_{i}\}_{i=1}^{m}$ and $n$ items $\{v_{j}\}_{j=1}^{n}$, the matrix factorization models map both users and items into a shared latent space with a low dimensionality $d \ll \min(m, n)$. For each user $u_{i}$, her latent features are represented by a latent vector $U_{i} \in \mathbb{R}^{1 \times d}$. Similarly, the latent features of the item $v_{j}$ are described by a latent vector $V_{j} \in \mathbb{R}^{1 \times d}$. In the PMF model, users' ratings on items are assumed to follow a Gaussian distribution as follows:
\begin{equation}
  p(R|U, V, \sigma^{2})=\prod_{i=1}^{m}\prod_{j=1}^{n}\left[\N(R_{ij}|
  U_{i}V_{j}^{\top}, \sigma^{2})\right]^{W_{ij}},
  \label{eq:pmf_gaussian}
\end{equation}
where $R \in \mathbb{R}^{m \times n}$ is the matrix denoting users' ratings on items; $U \in \mathbb{R}^{m \times d}$ and $V \in \mathbb{R}^{n \times d}$denote the latent features of all users and items, respectively; $\sigma^{2}$ is the variance of the Gaussian distribution; $W_{ij}$ is an indicator variable. If $u_{i}$ has rated $v_{j}$, $W_{ij}=1$, otherwise, $W_{ij}=0$. In addition, we also place zero-mean spherical Gaussian priors on $U$ and $V$ as:
\begin{equation}
 p(U|\sigma_{u}^{2})= \prod_{i=1}^{m}\N(U_{i}|0, \sigma_{u}^{2}I),~~
 p(V|\sigma_{v}^{2})= \prod_{j=1}^{n}\N(V_{j}|0, \sigma_{v}^{2}I).
 \label{eq:pmf1}
\end{equation}
where $I$ is the identity matrix, $\sigma_{u}^{2}$ and $\sigma_{v}^{2}$ are the variance parameters. Through the Bayesian inference, we have
\begin{equation}
  p(U, V|R, \sigma^{2}, \sigma_{u}^{2}, \sigma_{v}^{2})\propto
  p(R|U, V, \sigma^{2})p(U|\sigma_{u}^{2})p(V|\sigma_{v}^{2}).
 \label{eq:pmf2}
\end{equation}
The model parameters (\ie $U$ and $V$) can be learned via maximizing the log-posterior in Eq.~\eqref{eq:pmf2}, which is equivalent to solving the following problem:
\begin{equation}\label{eq:pmf3}
  \min_{U, V} \frac{1}{2} \|W \odot (R-UV^{\top})\|^{2}_{F}
  +\frac{\lambda_{u}}{2}\|U\|_{F}^{2}+\frac{\lambda_{v}}{2}\|V\|_{F}^{2},
\end{equation}
where $W\in \mathbb{R}^{m\times n}$ is the indicator matrix, and $W_{ij}$ is the $(i,j)$ element of $W$. In Eq.~\eqref{eq:pmf3}, $\odot$ denotes the Hadamard product of two matrices, $\lambda_{u}=\sigma^{2}/\sigma_{u}^{2}$, $\lambda_{v}=\sigma^{2}/\sigma_{v}^{2}$, and $\|\cdot\|_{F}$ denotes the Frobenius norm of a matrix.

\subsection{Social Recommendation Methods}
In recent years, lots of social recommendation approaches have been proposed~\cite{yang2014survey}. The Social Regularization (SR)~\cite{ma2011recommender} is one of the most representative methods. The SR method was implemented by matrix factorization framework. The basic idea was that a user may have similar interests with her social friends, and thus the learned latent features of a user and that of her social friends should be similar. The user similarity can be calculated using the Cosine similarity and Pearson correlation coefficient~\cite{su2009survey}.
To improve the performance of SR, \cite{yu2011adaptive} proposed an adaptive social similarity function. Moreover, as a user tends to trust different subsets of her social friends regarding with different domains, \cite{yang2012circle} introduced the circle-based recommendation (CircleCon) models, which considered domain-specific trust circles for recommendation. \cite{tang2013exploiting} exploited both the local and global social context for recommendation. \cite{ma2013experimental} extended the SR model to exploit users' implicit social relationships for recommendation. The implicit social relationships were defined between a user and other users that had most similar or dissimilar rating behaviors with her. In addition, \cite{li2015overlapping} extended SR to exploit the community structure of users' social networks to improve the recommendation accuracy. \cite{guo2015trustsvd} developed the TrustSVD model, which extended the SVD++ model~\cite{koren2008factorization} to incorporate the explicit and implicit influences from rated items and trusted users for recommendation. \cite{hu2015synthetic} proposed a recommendation framework named MR3, which jointly modeled users' rating behaviors, social relationships, and review comments.

\section{The Proposed Recommendation model}
\label{section:model}
This section presents the details of the proposed PRMF model that extends PMF to jointly learn the user preferences and the social dependency between users.

\subsection{Probabilistic Relational Matrix Factorization}
In Section~\ref{ssection:pmf}, the PMF model assumes the users are independent of each other (see~Eq.\eqref{eq:pmf1}). Thus, it ignores the social dependency between users. However, in practice, users are usually connected with each other through different types of social relationships, \eg trust relationships and online social relationships. In this work, to exploit users' social dependency for recommendation, we place the following priors on the user latent features:
\begin{equation}
  p(U|\sigma_{u}^{2}) \propto \big(\prod_{i=1}^{m}\mathcal{N}(U_{i}|0, \sigma_{u}^{2}I)\big)q(U),
  \label{eq:prmf1}
\end{equation}
where the first term of the priors is used to penalize the complexity of the latent features of each user, and the second term $q(U)$ is used to model the dependency between different users. Specifically, we define
\begin{equation}
q(U)=\mathcal{MN}_{m\times d}\big(0, \Sig^{-1}, I\big), ~~s.t.~~\Sig \succ 0,
\label{eq:multinormal}
\end{equation}
where $\mathcal{MN}_{a \times b}(M, A, B)$ denotes the matrix variate normal (MVN) distribution\footnote{
The density function for a random matrix $X$ following the MVN distribution $\mathcal{MN}_{a \times b}(M, A, B)$ is
\begin{equation}
  p(X)=\frac{\exp(-\frac{1}{2}\mbox{tr}\big[B^{-1}(X-M)^{\top}A^{-1}(X-M)\big])}{(2\pi)^{ab/2}|B|^{a/2}|A|^{b/2}}.\nonumber
\end{equation}
}
with mean $M \in \mathbb{R}^{a \times b}$, row covariance $A \in \mathbb{R}^{a \times a}$, and column covariance $B \in \mathbb{R}^{b \times b}$; $\Sig \succ 0$ indicates $\Sig$ is a positive definite matrix. In Eq.~\eqref{eq:multinormal}, $\Sig$ is the row precision matrix (\ie the inverse of the row covariance matrix) that models the relationships between different rows of $U$. In other words, $\Sig$ describes the social dependency between different users. Thus, $\Sig$ is called the social dependency matrix. In this work, for simplicity, we set the column covariance matrix of the MVN distribution as $I$, which indicates the user latent features in different dimensions are independent. Moreover, we can also rewrite Eq.~\eqref{eq:prmf1} as follows:
\begin{equation}
  p(U|\Sig, \sigma_{u}^{2})=\mathcal{MN}_{m\times d}\big(0, (\Sig+ \frac{1}{\sigma_{u}^{2}} I)^{-1}, I\big).
  \label{eq:mvn2}
\end{equation}

In practice, a user is usually only correlated with a small fraction of other users. Thus, it is reasonable to assume the social dependency matrix $\Sig$ is sparse. To achieve this objective, we introduce a sparsity-inducing prior for $\Sig$ as follows:
\begin{equation}
p(\Sig|\gamma) \propto \exp(-\frac{\gamma}{2} \|\Sig\|_{1}),
\end{equation}
where $\gamma$ is a positive constant used to control the sparsity of $\Sig$, and $\|\cdot\|_{1}$ is the $\ell_{1}$-norm of a matrix. In addition, the sparsity of $\Sig$ can also help improve the computation efficiency of the proposed model (see the discussions in Section~\ref{ssection:optimization}).

Moreover, we also assume users' ratings follow the Gaussian distribution in Eq.~\eqref{eq:pmf_gaussian}, and add Gaussian priors on the item latent features as in Eq.~\eqref{eq:prmf1}. Through the Bayesian inference, we have
\begin{align}
  &p(U, V, \Sig|R, \sigma^{2}, \sigma_{u}^{2}, \sigma_{v}^{2}, \gamma)\nonumber\\
 \propto&p(R|U, V, \sigma^{2})p(U|\Sig, \sigma_{u}^{2})p(\Sig|\gamma)p(V|\sigma_{v}^{2}).
 \label{eq:map2}
\end{align}
Then, the model parameters (\ie $U$, $V$, and $\Sig$) can be obtained by solving the following problem:
\begin{gather}
  \min_{U, V, \Sig \succ 0} \frac{1}{2\sigma^{2}} \|W \odot (R-UV^{\top})\|^{2}_{F}
  +\frac{1}{2\sigma_{u}^{2}}\|U\|_{F}^{2} +\frac{1}{2\sigma_{v}^{2}} \|V\|_{F}^{2}
  \nonumber\\
  +\frac{1}{2}\big[\mbox{tr}(U^{\top}\Sig U)
  -d\log |\Sig+\frac{1}{\sigma_{u}^{2}}I| + \gamma\|\Sig\|_{1}\big],
 \label{eq:obj1}
\end{gather}
where $|\cdot|$ denotes the determinant of a matrix. In Eq.~\eqref{eq:obj1}, the model parameters are learned without prior information about users' social relationships.

\subsubsection{Incorporating Prior Social Information}
Previous studies have demonstrated that user' explicit social relationships~\cite{yang2014survey} and implicit social relationships~\cite{ma2013experimental} can help improve the recommendation accuracy. These prior social information (\ie explicit and implicit social relationships) contains prior knowledge about users' social dependency in a recommendation task. Let $p_{0}(U)$ denote the MVN distribution of the user latent features $U$ derived from the prior information as follows:
\begin{equation}
  p_{0}(U) = \mathcal{MN}_{m\times d}\big(0, \Sigma, I\big),
\end{equation}
where $\Sigma \in \mathbb{R}^{m\times m}$ is the \emph{prior row covariance matrix}. If users' explicit social relationships are available, the elements of $\Sigma$ can be defined as follows:
\begin{eqnarray}
\small
\Sigma_{ik} = \left\{ \begin{array}{rl}
  cov(R_{i\ast}, R_{k\ast}) &\mbox{if $u_{k} \in \mathcal{F}(u_{i})$ or $u_{i}\in \mathcal{F}(u_{k})$ or i=k},\nonumber\\
  0 &\mbox{ otherwise},
       \end{array} \right.
\end{eqnarray}
where $R_{i\ast}$ is the $i^{th}$ row in $R$; $cov(x_{1}, x_{2})$ denotes the covariance between two observations; $\mathcal{F}(u_{i})$ denotes the set of $u_{i}$'s social friends. Moreover, if users' explicit social relationships are unavailable, we define the elements of $\Sigma$ as follows:
\begin{equation}
  \Sigma_{ik} = cov(R_{i\ast}, R_{k\ast}).
\end{equation}

To exploit the prior information for recommendation, we aims to minimize the distance between the learned MVN distribution $p(U|\Sig, \sigma_{u}^{2})$ and the MVN distribution $p_{0}(U)$ derived from prior information. This is achieved by minimizing the differential relative entropy between $p(U|\Sig, \sigma_{u}^{2})$ and $p_{0}(U)$ as follows:
\begin{align}
   &\min_{\Sig \succ 0} \int p_{0}(U)\log\frac{p_{0}(U)}{p(U|\Sig, \sigma_{u}^{2})}\nonumber\\
   &=\frac{d}{2}\mbox{tr}\big[(\Sig+\frac{1}{\sigma_{u}^{2}}I)\Sigma\big]
   -\frac{d}{2}\log\big|(\Sig+\frac{1}{\sigma_{u}^{2}}I)\Sigma\big|-\frac{md}{2}\nonumber\\
   &\propto\frac{1}{2}\mbox{tr}\big(\Sig \Sigma \big)-\frac{1}{2}\log|(\Sig+\frac{1}{\sigma_{u}^{2}}I)|.
   \label{eq:obj2}
\end{align}
The differential relative entropy in Eq.~\eqref{eq:obj2} can be obtained following the ideas in~\cite{dhillon2007differential}.

\subsubsection{A Unified Model}
Considering both the constraints in Eq.~\eqref{eq:obj1} and Eq.~\eqref{eq:obj2}, we formulate the final objective function of the proposed PRMF model as follows:
\begin{gather}
  \min_{U, V, \Sig \succ 0} \frac{1}{2} \|W \odot (R-UV^{\top})\|^{2}_{F}
  +\frac{\lambda_{u}}{2} \|U\|_{F}^{2}+\frac{\lambda_{v}}{2} \|V\|_{F}^{2}\nonumber\\
  +\frac{\alpha}{2}\big\{\mbox{tr}\big[\Sig(UU^{\top}+\beta\Sigma) \big]
  -(d+\beta)\log \big|\Sig+\frac{\lambda_{u}}{\alpha}I\big| + \gamma\|\Sig\|_{1} \big\},
\label{eq:obj4}
\end{gather}
where $\lambda_{u}=\sigma^{2}/\sigma_{u}^{2}$, $\lambda_{v}=\sigma^{2}/\sigma_{v}^{2}$, and $\alpha=\sigma^{2}$; $\beta$ is the parameter that controls the contribution from prior information. 

\subsection{Optimization Algorithm}
\label{ssection:optimization}
The optimization problem in Eq.~\eqref{eq:obj4} can be solved by using the following alternating algorithm.

\noindent \textbf{Optimize} $U$ and $V$: By fixing $\Sig$, the optimization problem in Eq.~\eqref{eq:obj4} becomes:
\begin{gather}
\min_{U, V}\frac{1}{2} \|W \odot (R-UV^{\top})\|^{2}_{F}
  +\frac{\lambda_{u}}{2} \|U\|^{2}_{F}+\frac{\lambda_{v}}{2} \|V\|^{2}_{F}\nonumber\\
  +\frac{\alpha}{2}\mbox{tr}(U^{\top}\Sig U).
\label{eq:subopt1}
\end{gather}
The optimization problem in Eq.~\eqref{eq:subopt1} can be solved using the SGD algorithm~\cite{koren2009matrix}. The updating rules used to learn the latent features are as follows:
\begin{eqnarray}
  U_{i} &\leftarrow& U_{i} +\theta \big(\Delta_{ij}V_{j}-\lambda_{u}U_{i}-\alpha \Sig_{i\ast}U\big)\nonumber\\
  V_{j} &\leftarrow& V_{j} + \theta \big(\Delta_{ij}U_{i}-\lambda_{v}V_{j}\big),
  \label{eq:sgd}
\end{eqnarray}
where $\Delta_{ij}=R_{ij}-U_{i}V_{j}^{\top}$, $\Sig_{i\ast}$ is the $i^{th}$ row of $\Sig$, and $\theta$ is the learning rate. Note the SGD updates are only performed on the observed rating pairs $\mathcal{D}=\{(u_{i}, v_{j}, R_{ij})|W_{ij}>0\}$.

\noindent \textbf{Optimize} $\Sig$: By fixing $U$ and $V$, the optimization problem with respect to $\Sig$ is as follows:
\begin{equation}
\min_{\Sig \succ 0} \mbox{tr}\big[\Sig(UU^{\top}+\beta\Sigma)\big]
-(d+\beta)\
\log |\Sig+\frac{\lambda_{u}}{\alpha}I|+\gamma\|\Sig\|_{1}.
\label{eq:subopt2}
\end{equation}
The optimization problem in Eq.~\eqref{eq:subopt2} is convex with respect to $\Sig$. Therefore, the optimal solution $\widehat{\Sig}$ to Eq.~\eqref{eq:subopt2} satisfies:
\begin{equation}
  (\widehat{\Sig}+\frac{\lambda_{u}}{\alpha}I)^{-1} - \frac{1}{d+\beta}(UU^{\top}+\beta\Sigma)=\frac{\gamma}{d+\beta}\widehat{G},
  \label{eq:subopt3}
\end{equation}
where $\widehat{G}$ is the sub-gradient, and the elements of $\widehat{G}$ are in $[-1, 1]$. Therefore, following the derivation of Dantzig estimator~\cite{candes2007dantzig}, we drop the constraint $\Sig \succ 0$ and consider the following optimization problem:
\begin{gather}
 \min \|\Sig\|_{1}~~~ s.t. \nonumber\\
 \big\|(\Sig+\frac{\lambda_{u}}{\alpha}I)^{-1} - \frac{1}{d+\beta}(UU^{\top}+\beta\Sigma)\big\|_{\infty}\leq \frac{\gamma}{d+\beta}.
 \label{eq:subopt4}
\end{gather}
By multiplying $\Sig+\frac{\lambda_{u}}{\alpha}I$ with the constraint, we obtain the following relaxation of Eq.~\eqref{eq:subopt4}:
\begin{gather}
 \min \|\Sig\|_{1} ~~s.t.~~
 \big\|C\Sig - E \big\|_{\infty}\leq \tau,
 \label{eq:subopt5}
\end{gather}
where $C=\frac{1}{d+\beta}(UU^{\top}+\beta\Sigma)$, $E=I-\frac{\lambda_{u}}{\alpha}C$, and $\tau=\frac{\gamma}{d+\beta}$. This relaxation has been used in the constrained $\ell_{1}$-minimization for inverse matrix estimation (CLIME)~\cite{cai2011constrained}. Indeed, the optimization in Eq.~\eqref{eq:subopt5} is equivalent to the following optimization problem:
\begin{equation}
  \min_{\Sig}\mbox{tr}\big(\Sig^{\top}C \Sig\big)-\mbox{tr}(E\Sig) + \tau \|\Sig\|_{1}.
  \label{eq:subopt6}
\end{equation}
Let $\widetilde{\Sig}$ be the solution of Eq.~\eqref{eq:subopt6}, which is not necessarily symmetric. We use the following symmetrization step to obtain the final $\widehat{\Sig}$,
\begin{equation}\label{eq:solution}
  \widehat{\Sig}_{ik}=\widehat{\Sig}_{ki}=\widetilde{\Sig}_{ik}\mathbb{I}(|\widetilde{\Sig}_{ik}|\leq|\widetilde{\Sig}_{ki}|)
  +\widetilde{\Sig}_{ki}\mathbb{I}(|\widetilde{\Sig}_{ik}|>|\widetilde{\Sig}_{ki}|),
\end{equation}
where $\mathbb{I}(\cdot)$ is an indicator function, which
is equal to 1 if the condition is satisfied, and otherwise 0. In addition, although constraint $\Sig \succ 0$ is not imposed in Eq.~\eqref{eq:subopt6}, the symmetrized $\widehat{\Sig}$ is positive definite with high probability and converge to the solution of Eq.~\eqref{eq:subopt2} under the spectral norm, guaranteed by Remark 1 and Theorem 1 in~\cite{liu2012high}.

\begin{algorithm}[t]\label{algorithm:prml}
  \caption{PRMF Optimization Algorithm}
  \LinesNumbered
  \SetKwInOut{KwInput}{Input}
  \SetKwInOut{KwOutput}{Output}

  \BlankLine
  \KwInput{$\mathcal{D}$, $\Sigma$, $d$, $\lambda_{u}$, $\lambda_{v}$, $\alpha$, $\beta$, $\gamma$, $\theta$, $\rho$}
  \KwOutput{$U$, $V$, $\Sig$}
  \If{$\beta>0$}{
    $X \leftarrow SVD(\Sigma, d)$;
  }
  Initialize $U$ and $V$ randomly, and set $\Sig=I$; \\
  \For{$iter=1, 2, \dots, max\_iter$}{
      \For{$t=1, 2, \dots, T$}{
          \ForEach{$(u_{i}, v_{j}, R_{ij})\in \mathcal{D}$}{
          $U_{i} \leftarrow U_{i} +\theta \big(\Delta_{ij}V_{j}-\lambda_{u}U_{i}-\alpha \Sig_{i\ast}U\big)$;\\
          $V_{j} \leftarrow V_{j} + \theta \big(\Delta_{ij}U_{i}-\lambda_{v}V_{j}\big)$;\\
          }
      }
      \If{$\beta=0$}{$\widehat{U}=\frac{1}{\sqrt{d}}U$; $\tau=\frac{\gamma}{d}$; }
       \Else{$\widehat{U} = [\frac{1}{\sqrt{d+\beta}}U, \frac{\sqrt{\beta}}{\sqrt{d+\beta}}X]$; $\tau=\frac{\gamma}{d+\beta}$;}
       $Z^{0}=\Sig$; $Y^{0} = 0$; $E=I-\frac{\lambda_{u}}{\alpha}\widehat{U}\widehat{U}^{\top}$;\\
       $P=I-\frac{1}{\rho}\widehat{U}(I+\frac{1}{\rho}\widehat{U}^{\top}\widehat{U})^{-1}\widehat{U}^{\top}$;\\
      \For{$t=0, 1, \dots, K-1$}{
       $\Sig^{t+1} = soft\big[Z^{t}-Y^{t}, \frac{\tau}{\rho}\big]$;\\
       $Z^{t+1} = P(\frac{1}{\rho}E+\Sig^{t+1}+Y^{t})$;\\
      $Y^{t+1} = Y^{t}+\Sig^{t+1}-Z^{t+1}$;
      }
      $\Sig = \Sig^{K}$;
  }
\end{algorithm}

The optimization problem in Eq.~\eqref{eq:subopt6} can be solved using the ADMM algorithm~\cite{boyd2011distributed}. The augmented Lagrangian of Eq.~\eqref{eq:subopt6} is as follows:
\begin{align}
  L_{\rho}(\Sig, Z, Y)=&\mbox{tr}\big(Z^{\top}C Z\big)-\mbox{tr}(EZ) + \tau \|\Sig\|_{1}+\langle Y, \Sig-Z\rangle \nonumber\\
  &+\frac{\rho}{2}\|\Sig-Z\|_{F}^{2},
\end{align}
where $Y$ is a scaled dual variable and $\rho>0$. We can obtain the following ADMM updates:
\begin{align}
  \Sig^{t+1} &= \argmin \frac{\rho}{2}\|\Sig-Z^{t}+Y^{t}\|_{F}^{2}+\tau \|\Sig\|_{1}, \label{eq:admm_om}\\
  Z^{t+1} &= \argmin\frac{\rho}{2}\|\Sig^{t+1}-Z+Y^{t}\|_{F}^{2}+\mbox{tr}(Z^{\top}C Z)-\mbox{tr}(EZ), \label{eq:admm_z}\\
  Y^{t+1} &= Y^{t}+\Sig^{t+1}-Z^{t+1}.\label{eq:admm_y}
\end{align}
The closed solution to Eq.~\eqref{eq:admm_om} is as follows:
\begin{equation}
  \Sig^{t+1} = soft\big[Z-Y, \frac{\tau}{\rho}\big],
\end{equation}
where
\begin{eqnarray}
soft\big[A, \lambda\big] = \left\{ \begin{array}{rl}
  A_{ij}-\lambda &\mbox{if $A_{ij}>\lambda$,} \\
  A_{ij}+\lambda &\mbox{if $A_{ij}<-\lambda$,}\\
  0 &\mbox{ otherwise}.
       \end{array} \right.
\end{eqnarray}
The solution to Eq.~\eqref{eq:admm_z} is as follows:
\begin{equation}
  Z = (\frac{1}{\rho}C + I)^{-1}(\frac{1}{\rho}E+\Sig+Y).
  \label{eq:admm_z2}
\end{equation}
The time complexity of the inverse operation in Eq.~\eqref{eq:admm_z2} is $\mathcal{O}(m^{3})$. As $\Sigma$ is symmetric, to improve the computation efficiency, we consider the following approximation $\Sigma\approx XX^{\top}$, where $X\in \mathbb{R}^{m \times d}$. In this paper, we obtain $X$ by factorizing $\Sigma$ using SVD and choosing the singular vectors with respect to the top $d$ largest singular values (see line 2 in Algorithm~\ref{algorithm:prml}). Let $\widehat{U} = [\frac{1}{\sqrt{d+\beta}}U, \frac{\sqrt{\beta}}{\sqrt{d+\beta}}X] \in \mathbb{R}^{m \times 2d}$. When the prior covariance matrix $\Sigma$ is not available, we set $\widehat{U} = \frac{1}{\sqrt{d}}U$. Then, $C\approx \widehat{U}\widehat{U}^{\top}$. Following Woodbury matrix identity, we have
\begin{equation}
  (\frac{1}{\rho}C + I)^{-1} \approx I-\frac{1}{\rho}\widehat{U}(I+\frac{1}{ \rho}\widehat{U}^{\top}\widehat{U})^{-1}\widehat{U}^{\top}.
\end{equation}
Using this matrix operation, the time complexity of the update of $Z$ in Eq.~\eqref{eq:admm_z2} becomes $\mathcal{O}(dm^{2})$. The details of the proposed optimization algorithm are summarized in Algorithm~\ref{algorithm:prml}. At each iteration, the time complexity of the SGD updates is $\mathcal{O}(T \cdot |\mathcal{D}| \cdot \bar{m} \cdot d)$, where $\bar{m}$ denotes the average number of nonzero elements in each row of $\Sig$. Thus, the sparsity of $\Sig$ can help improve the computation efficiency of the proposed method. The time complexity of the ADMM updates is $\mathcal{O}(K \cdot d \cdot m^{2})$. Empirically, we set $T=30$ and $K=30$ in our experiments.

\section{Experiments}
\label{section:experiments}
In this section, we conduct empirical experiments on real datasets to demonstrate the effectiveness of the proposed recommendation method.

\subsection{Experimental Setting}

\subsubsection{Dataset Description}
The experiments are performed on four public datasets: MovieLens-100K\footnote{http://grouplens.org/datasets/movielens/100k/}, MovieLens-1M\footnote{http://grouplens.org/datasets/movielens/1m/}, Ciao, and Epinions\footnote{http://www.public.asu.edu/~jtang20/datasetcode/truststudy.htm}. MovieLens-100K contains 100,000 ratings given by 943 users to 1,682 movies. MovieLens-1M consists of 1,000,209 ratings given by 6,040 users to 3,706 movies. We denote these two datasets by ML-100K and ML-1M, respectively. For the Ciao and Epinions datasets, we remove the items that have less than $3$ ratings. Finally, on Ciao dataset, we have 185,042 ratings given by 7,340 users to 21,881 items. On Epinions dataset, there are 642,104 ratings given by 22,112 users to 59,104 items. The densities of the rating matrices on these datasets are 6.30\% for ML-100K, 4.47\% for ML-1M, 0.12\% for Ciao, and 0.05\% for Epinions. Moreover, on Ciao and Epinions datasets, we have observed 111,527 and 353,419 social relationships between users. The densities of the social relation matrices are 0.21\% for Ciao and 0.07\% for Epinions. Table~\ref{tab:dataset} summarizes the details of the experimental datasets.

For each dataset, $80\%$ of the observed ratings are randomly sampled for training, and the remaining $20\%$ observed ratings are used for testing.

\subsubsection{Evaluation Metrics}
The performances of the recommendation algorithms are evaluated by two most popular metrics: mean absolute error (MAE) and root mean square error (RMSE). The definitions of MAE and RMSE are as follows:
\begin{eqnarray}
  MAE&=& \frac{1}{|\mathcal{D}_{test}|}\sum_{(u_{i}, v_{j})\in \mathcal{D}_{test}} |R_{ij}-\hat{R}_{ij}|,\\
  RMSE&=&\sqrt{\frac{1}{|\mathcal{D}_{test}|}\sum_{(u_{i}, v_{j})\in \mathcal{D}_{test}}(R_{ij}-\hat{R}_{ij})^{2}},
\end{eqnarray}
where $R_{ij}$ denotes the observed rating in the testing data, $\hat{R}_{ij}$ is the predicted rating, and $|\mathcal{D}_{test}|$ is the number of tested ratings. From the definitions, we can notice that lower MAE and RMSE values indicate better recommendation accuracy.

\subsubsection{Evaluated Recommendation Methods}
We compare the following recommendation methods: (1) \emph{\textbf{PMF}}: This is the probabilistic matrix factorization model introduced in Section~\ref{ssection:pmf}; (2) $\mbox{\emph{\textbf{SR}}}^{imp}$: This is the SR method that exploits users' implicit social relationships for recommendation~\cite{ma2013experimental}; (3) $\mbox{\emph{\textbf{SR}}}^{exp}$: This is the SR method that exploits users' explicit social relationships for recommendation~\cite{ma2011recommender}; (4) \emph{\textbf{LOCABAL}}: This is the social recommendation model proposed in~\cite{tang2013exploiting}, which exploits both local and global social context for recommendation; (5) \emph{\textbf{eSMF}}: This is the extended social matrix factorization model that extends LOCABL to consider the graph structure of social neighbors for recommendation~\cite{hu2015synthetic}; (6) \emph{\textbf{PRMF}}: This method learns users' social dependency without prior information about users' social relationships. The objective function is Eq.~\eqref{eq:obj1}; (7) $\mbox{\emph{\textbf{PRMF}}}^{imp}$: This is the proposed method that exploits users' implicit social relationships as the prior knowledge used to learn users' social dependency; (8) $\mbox{\emph{\textbf{PRMF}}}^{exp}$: This is the proposed method that exploits users' explicit social relationships to learn the social dependency between users.

\begin{table}
\centering
\small
\caption{The statistics of the experimental datasets.}
\label{tab:dataset}
\begin{tabular}{l|cc|cc} \hline
    & ML-100K & ML-1M & Ciao & Epinions\\\hline\hline
\# Users &  943 & 6,040 & 7,340  & 22,112\\
\# Items & 1682 & 3,706 & 21,881 & 59,104\\
\# Ratings & 100,000 & 1,000,209 & 185,042 & 642,104 \\
Rating Density & 6.30\% & 4.47\% &0.12\% & 0.05\%\\
\hline
\# Social Rel. & N.A. & N.A. & 111,527 & 353,419\\
Social density & N.A. & N.A. & 0.21\% & 0.07\%\\
\hline
\end{tabular}
\vspace{-12px}
\end{table}

\begin{figure*}
  \centering
  \subfigure[ML-100K]{
  \includegraphics[width=3in]{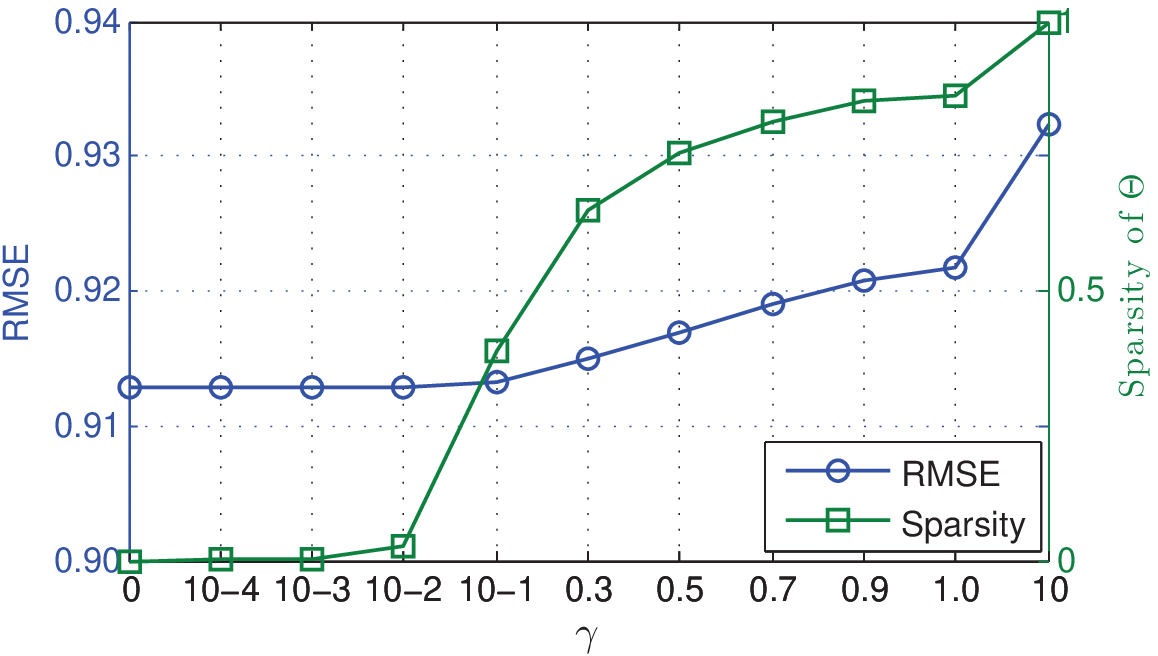}
  \label{fig:ml100k}
  }
  \subfigure[Ciao]{
  \includegraphics[width=3in]{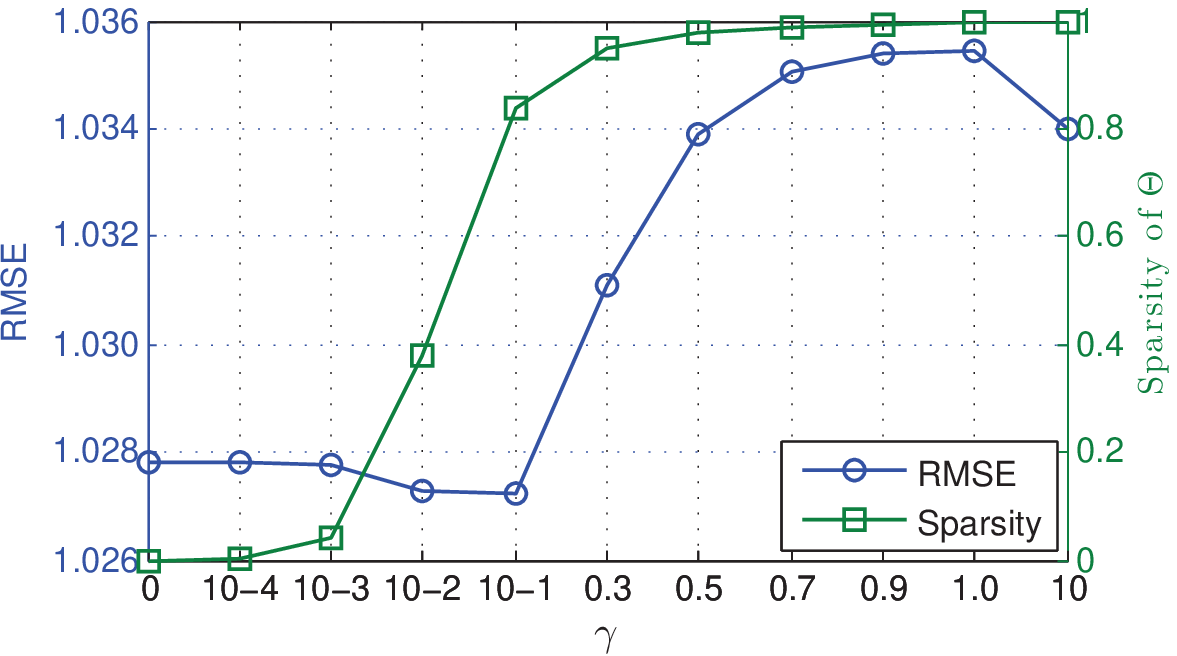}
  \label{fig:ciao}
  }
  \vspace{-8px}
  \caption{Performance trend of $\mbox{PRMF}^{imp}$ on the ML-100K and Ciao datasets measured by RMSE with different settings of $\gamma$.}
  \vspace{-12px}
  \label{fig:sparse}
\end{figure*}

\begin{table}
  \caption{Performance comparisons on datasets without users' explicit social relationships.}
  \centering
  \small
  \label{tab:results}
  \begin{tabular}{l|l|c|c}
  \hline
  \multirow{2}{*}{Dataset} & \multirow{2}{*}{Method} & \multirow{2}{*}{RMSE} & \multirow{2}{*}{MAE} \\
  & & & \\ \hline \hline
  \multirow{4}{*}{ML-100K}
  & PMF & 0.9251$\pm$0.0021 & 0.7314$\pm$0.0017  \\ \cline{2-4}
  & $\mbox{SR}^{imp}$ & 0.9205$\pm$0.0013 & 0.7290$\pm$0.0008 \\ \cline{2-4}
  & PRMF & 0.9157$\pm$0.0006 & 0.7226$\pm$0.0005 \\ \cline{2-4}
  & $\mbox{PRMF}^{imp}$ & \textbf{0.9132$\pm$0.0003} & \textbf{0.7210$\pm$0.0005} \\
  \hline \hline
    \multirow{4}{*}{ML-1M}
  & PMF & 0.8728$\pm$0.0029 & 0.6807$\pm$0.0023   \\ \cline{2-4}
  & $\mbox{SR}^{imp}$ & 0.8621$\pm$0.0008 & 0.6739$\pm$0.0006 \\ \cline{2-4}
  & PRMF & 0.8592$\pm$0.0009 & 0.6738$\pm$0.0010 \\ \cline{2-4}
  & $\mbox{PRMF}^{imp}$ & \textbf{0.8572$\pm$0.0009}&\textbf{0.6727$\pm$0.0010} \\
  \hline
\end{tabular}
\vspace{-11px}
\end{table}

\begin{table}
  \caption{Performance comparisons on datasets with users' explicit social relationships.}
  \centering
  \small
  \label{tab:resultsprio}
  \begin{tabular}{l|l|c|c}
  \hline
  \multirow{2}{*}{Dataset} & \multirow{2}{*}{Method} & \multirow{2}{*}{RMSE} & \multirow{2}{*}{MAE} \\
  & & & \\ \hline \hline
  \multirow{8}{*}{Ciao}
  & PMF & 1.1031$\pm$0.0045 & 0.8439$\pm$0.0034  \\ \cline{2-4}
  & $\mbox{SR}^{imp}$ & 1.0597$\pm$0.0042 &0.8234$\pm$0.0036 \\ \cline{2-4}
  & $\mbox{SR}^{exp}$ & 1.0286$\pm$0.0029 & \textbf{0.7951$\pm$0.0020} \\ \cline{2-4}
  & LOCABAL & 1.0777$\pm$0.0039 & 0.8330$\pm$0.0022 \\ \cline{2-4}
  & eSMF & 1.0592$\pm$0.0043 & 0.8122$\pm$0.0010 \\ \cline{2-4}
  & PRMF & 1.0326$\pm$0.0030 & 0.8006$\pm$0.0023 \\ \cline{2-4}
  & $\mbox{PRMF}^{imp}$ & 1.0305$\pm$0.0030 & 0.7993$\pm$0.0023 \\ \cline{2-4}
  & $\mbox{PRMF}^{exp}$ & \textbf{1.0258$\pm$0.0035} & 0.7980$\pm$0.0025 \\
  \hline \hline
    \multirow{8}{*}{Epinions}
  & PMF & 1.1613$\pm$0.0022 & 0.8932$\pm$0.0019   \\ \cline{2-4}
  & $\mbox{SR}^{imp}$ & 1.1321$\pm$0.0054 & 0.8874$\pm$0.0016 \\ \cline{2-4}
  & $\mbox{SR}^{exp}$ & 1.1263$\pm$0.0027 & 0.8795$\pm$0.0023 \\ \cline{2-4}
  & LOCABAL & 1.1289$\pm$0.0008 & 0.8734$\pm$0.0010 \\ \cline{2-4}
  & eSMF & 1.1306$\pm$0.0025 & 0.8749$\pm$0.0025 \\ \cline{2-4}
  & PRMF & 1.1097$\pm$0.0024 & 0.8703$\pm$0.0022 \\ \cline{2-4}
  & $\mbox{PRMF}^{imp}$ & \textbf{1.1081$\pm$0.0023} & \textbf{0.8691$\pm$0.0022} \\ \cline{2-4}
  & $\mbox{PRMF}^{exp}$ & 1.1085$\pm$0.0024 & 0.8695$\pm$0.0022\\
  \hline
\end{tabular}
\vspace{-10px}
\end{table}

\subsubsection{Parameter Settings}
We adopt cross-validation to choose the parameters for the evaluated algorithms. The validation data is constructed by randomly chosen 10\% of the ratings in the training data. For matrix factorization methods, we set the dimensionality of the latent space $d$ to 10. The latent features of users and items are randomly initialized by a Gaussian distribution with mean 0 and standard deviation $1/\sqrt{d}$. Moreover, we set the regularization parameters $\lambda_{u}=\lambda_{v}$ and choose the parameters from $\{10^{-5}, 10^{-4}, \cdots, 10^{-1}\}$. For PRMF, $\alpha$ is chosen from $\{2^{-5}, 2^{-4}, \cdots, 2^{-1}\}$, $\theta$ is chosen from $\{2^{-5}, 2^{-4}, \cdots, 2^{-1}\}$. Moreover, we set $\gamma=10^{-4}$, $\beta=10$, and $\rho=100$. For SR methods, the regularization parameter $\alpha$ is chosen from $\{2^{-7}, 2^{-6}, \cdots, 2^{0}\}$. The user similarity is computed using Pearson correlation coefficient, and we set the threshold of the user similarity at 0.75 and $N=10$, following~\cite{ma2013experimental}. The parameters of LOCABAL and eSMF are set following~\cite{tang2013exploiting} and~\cite{hu2015synthetic}.

\subsection{Summary of Experiments}
Table~\ref{tab:results} summarizes the experimental results on the datasets without users explicit social relationships, and Table~\ref{tab:resultsprio} summarizes the results on the datasets with users' explicit social relationships. We make the following observations:
\begin{itemize}
 \item On all datasets, PRMF outperforms PMF by 0.94\% on ML-100K, 1.36\% on ML-1M, 7.05\% on Ciao, and 5.16\% on Epinions, in terms of RMSE. This indicates the recommendation accuracy can be improved by jointly learning users' preferences and users' social dependency.
 \item Compared with $\mbox{SR}^{imp}$, $\mbox{PRMF}^{imp}$ achieves better results on all datasets. For example, in terms of RMSE, $\mbox{PRMF}^{imp}$ outperforms $\mbox{SR}^{imp}$ by 0.73\%, 0.49\%, 2.92\%, and 2.40\%, respectively. This shows that users' social dependency learned from the rating data is more effective than the user dependency predefined based on users' implicit social relationships, for improving the recommendation accuracy.
 \item The proposed $\mbox{PRMF}^{exp}$ method outperforms the state-of-the-art social recommendation methods that exploits users' explicit social relationships for recommendation. On the Ciao and Epinions datasets, the average improvements of $\mbox{PRMF}^{exp}$ over $\mbox{SR}^{exp}$, LOCABAL, and eSMF, in terms of RMSE, are 1.03\%, 3.60\%, 2.77\%, respectively. This again demonstrates the effectiveness of the proposed method.
 \item $\mbox{PRMF}^{imp}$ and $\mbox{PRMF}^{exp}$ outperforms PRMF on all datasets. This indicates the prior information (\eg users' explicit and implicit social relationships) are beneficial in improving recommendation accuracy. However, the improvements are not very significant. One potential reason is the SVD factorization used in Algorithm~\ref{algorithm:prml} (line 2) may not accurately approximate the prior covariance matrix.
\end{itemize}

In addition, we also study the impact of the sparsity of the learned social dependency matrix $\Sig$ on the recommendation accuracy. We choose the regularization parameter $\gamma$ from $\{0, 10^{-4}, 10^{-3}, 10^{-2}, 0.1, 0.3, 0.5, 0.7, 0.9, 1.0, 10\}$. Figure~\ref{fig:sparse} shows the performance trend of $\mbox{PRMF}^{imp}$ on the ML-100K and Ciao datasets, in terms of RMSE. Observed that the sparsity of $\Sig$ increases with the increase of $\gamma$. As shown in Figure~\ref{fig:ml100k}, denser social dependency matrix generally achieves better recommendation accuracy. However, denser social dependency matrix cause more computation time used to learn the user latent features. Indeed, there exists some balance between the computation efficiency and the recommendation accuracy. For example, on the ML-100K dataset, by setting $\gamma$ to $0.3$, the sparsity of the learned $\Sig$ is 65.21\%, and the RMSE value is 0.9149, which is 0.56\% better than the best competitor $\mbox{SR}^{imp}$. Moreover, Figure~\ref{fig:ciao} also indicates that better recommendation accuracy may be achieved by learning a sparse $\Sig$. For example, on the Ciao dataset, the best recommendation accuracy is achieved by setting $\gamma$ to 0.1, and the sparsity of the learned $\Sig$ is 79.74\%. On the ML-1M and Epinions datasets, we have similar observations with that on the ML-100K dataset. Due to space limitation, we do not report those results here.

\section{Conclusion and Future Work}
In this paper, we propose a novel social recommendation method, named probabilistic relational matrix factorization (PRMF). For a specific recommendation task, the proposed approach jointly learns users' preferences and the optimal social dependency between users, to improve the recommendation accuracy. Empirical results on real datasets demonstrate the effectiveness of PRMF, in comparison with start-of-the-art social recommendation algorithms.

The future work will focus on the following potential directions. First, we would like to develop more efficient optimization algorithms for PRMF, based on the parallel optimization framework proposed in~\cite{wang2013large}. Second, we are also interested in extending PRMF to solve the top-\emph{N} item recommendation problems.

\bibliographystyle{named}
\bibliography{sigproc}

\end{document}